\documentclass[aps,pra,twocolumn,showpacs,preprintnumbers]{revtex4-1}

\usepackage{psfrag,graphicx}
\usepackage{dcolumn}
\usepackage{amsmath,amssymb}
\usepackage{amsfonts}
\usepackage{bm}
\usepackage{amsfonts,amssymb,amsmath}
\usepackage{bbold}
\usepackage{epstopdf}
\usepackage{xcolor}
\usepackage{graphicx}
\usepackage{color}

\usepackage{verbatim}

\DeclareSymbolFont{bbold}{U}{bbold}{m}{n}
\DeclareSymbolFontAlphabet{\mathbbold}{bbold}

\usepackage[caption=false]{subfig}

\begin{document}
	
\title{Kramers pairs of Majorana corner states in a topological insulator bilayer}
\author{Katharina Laubscher, Danial Chughtai, Daniel Loss, and Jelena Klinovaja}
\affiliation{Department of Physics, University of Basel, Klingelbergstrasse 82, CH-4056 Basel, Switzerland}
	
\begin{abstract}	
We consider a system consisting of two tunnel-coupled two-dimensional topological insulators proximitized by a top and bottom superconductor with a phase difference of $\pi$ between them. We show that this system exhibits a time-reversal invariant second-order topological superconducting phase characterized by the presence of a Kramers pair of Majorana corner states at all four corners of a rectangular sample. We furthermore investigate the effect of a weak time-reversal symmetry breaking perturbation and show that an in-plane Zeeman field leads to an even richer phase diagram exhibiting two nonequivalent phases with two Majorana corner states per corner as well as an intermediate phase with only one Majorana corner state per corner. We derive our results analytically from continuum models describing our system. In addition, we also provide independent numerical confirmation of the resulting phases using discretized lattice representations of the models, which allows us to demonstrate the robustness of the topological phases and the Majorana corner states against parameter variations and potential disorder.
\end{abstract}
	
\maketitle
	
\section{Introduction}
	
Motivated by the seminal work on one-dimensional (1D) $p$-wave superconductors~\cite{Kitaev2001}, Majorana bound states have been predicted to occur in a variety of condensed matter systems as a signature of a topologically non-trivial superconducting phase. Apart from their fundamental interest, Majorana bound states are considered to be promising building blocks for topologically protected qubits due to their non-Abelian braiding statistics. Many well-known proposals for the experimental realization of Majorana bound states rely on the competition between a strong magnetic field and proximity-induced superconducting pairing~\cite{Lutchyn2018,Prada2019}. However, such setups suffer from the disadvantage that a strong magnetic field itself has a detrimental effect on superconductivity. To circumvent this issue, the concept of time-reversal invariant topological superconductivity has raised significant interest. In this case, Kramers pairs of Majorana bound states emerge in the absence of a magnetic field~\cite{Fu2008,Qi2009,Sato2009,Wong2012,Zhang2013,Keselman2013,Klinovaja2014,Haim2014,Hsu2018,Volpez2019,Schrade2015,Haim2016,Schrade2017,Gaidamaiskas2014,Klinovaja2014b,Ebisu2016,Thakurathi2018,Nakosai2013,Dumitrescu2014,Aligia2018}.

In the standard proposals, Majorana bound states are realized at the zero-dimensional edges of 1D topological superconductors (TSCs). More recently, the notion of topological insulators (TIs) and TSCs has been extended to capture also their \textit{higher-order} generalizations~\cite{Benalcazar2017,Benalcazar2017b,Geier2018,Peng2017,Schindler2018,Song2017,Imhof2017}. While conventional $d$-dimensional TIs and TSCs exhibit gapless edge states at their $(d-1)$-dimensional boundaries, $n$th-order $d$-dimensional TIs or TSCs exhibit gapless edge states at their $(d-n)$-dimensional boundaries. In particular, a two-dimensional (2D) second-order topological superconductor (SOTSC) hosts Majorana bound states at the corners of a rectangular sample. By now, a large variety of platforms hosting such Majorana corner states (MCSs) has been proposed. While most of these proposals use an applied magnetic field to induce the second-order phase~\cite{Volpez2018,Laubscher2019,Yan2019,Zhang2019a,Laubscher2020,Wu2019,Franca2019,Plekhanov2019,Ahn2020,Liu2018,Ghorashi2019,Ghorashi2020}, the case of time-reversal invariant SOTSCs with Kramers pairs of MCSs has been studied less extensively. The few setups proposed so far rely on unconventional superconductivity as the relevant mechanism driving the transition to the second-order phase~\cite{Wang2018, Yan2018}. This motivates us to look for an alternative model realizing a time-reversal invariant SOTSC based on conventional ingredients only. The setup we propose consists of two tunnel-coupled 2D TIs, each described by a Bernevig-Hughes-Zhang (BHZ) model, proximitized by a top and bottom superconductor of a phase difference of $\pi$, see Fig.~1. In the absence of interlayer tunneling and superconductivity, each TI layer hosts a pair of gapless helical edge states. Once interlayer tunneling and superconductivity are turned on, these edge states are gapped out. However, the resulting phase is not necessarily trivial. Indeed, we show that in a certain region of parameter space, the system is a SOTSC with a Kramers pair of MCSs at all four corners of a rectangular sample. These corner states are protected by particle-hole and time-reversal symmetry and cannot be removed unless one of the protecting symmetries is broken or the edge gap closes and reopens.

\begin{figure}[t!]
\centering
\includegraphics[width=0.9\columnwidth]{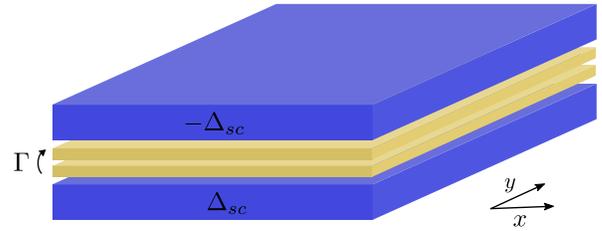}
\caption{The setup consists of two 2D TI layers (yellow) coupled by an interlayer tunneling term of strength $\Gamma$. The two layers are proximitized by a top and bottom superconductor (blue) with a phase difference of $\pi$ between them.}
\label{fig:model}
\end{figure} 

The paper is organized as follows. In Sec.~\ref{sec:model} we describe our setup, which consists of two tunnel-coupled 2D TIs, each described by a BHZ model, in proximity to a top and a bottom superconductor of a phase difference of $\pi$, see Fig.~\ref{fig:model}. In Sec.~\ref{sec:TI}, we obtain expressions for the gapless edge states appearing in the absence of superconductivity and interlayer tunneling. In this case, our model simply corresponds to two decoupled BHZ layers. In Sec.~\ref{sec:cornerstates}, we then perturbatively account for weak interlayer tunneling as well as weak proximity-induced superconductivity. As a consequence, the helical edge states found previously are gapped out. We show that there exists a regime of parameters for which the system is a time-reversal invariant SOTSC with a Kramers pair of MCSs at all four corners of a rectangular sample. In order to account for a possible complication in some experimentally relevant setups, we additionally comment on the case of unequal interlayer tunneling amplitudes for particle-like and hole-like bands in Sec.~\ref{sec:exp}. Finally, in Sec.~\ref{sec:single_MCS}, we discuss the case of broken time-reversal symmetry in the presence of a weak in-plane Zeeman field. We show that this enriches the phase diagram further, allowing us to access also a SOTSC phase with a single MCS per corner. We summarize our results in Sec.~\ref{sec:conclusions}.

\section{Model}
\label{sec:model}

We consider a 2D TI bilayer, where each of the two TI layers is described by a BHZ model~\cite{Bernevig2006}. In momentum space, the Hamiltonian of a single TI layer can then be written as $H_0=\sum_{\mathbf{k}}\Psi_\mathbf{k}^\dagger\mathcal{H}_{0}(\mathbf{k})\Psi_\mathbf{k}$ in the basis $\Psi_\mathbf{k}=(\psi_{\mathbf{k}11}$, $\psi_{\mathbf{k}1\bar1}$, $\psi_{\mathbf{k}\bar11}$, $\psi_{\mathbf{k}\bar1\bar1}$), where $\psi_{\mathbf{k}\sigma s}$ ($\psi_{\mathbf{k}\sigma s}^\dagger$) destroys (creates) an electron with in-plane momentum $\mathbf{k}=(k_x,k_y)$, 
orbital degree of freedom $\sigma\in\{1,\bar 1\}$ and spin $s\in\{1,\bar 1\}$. The Hamiltonian density is given by~\cite{comment}
\begin{align}
&\mathcal{H}_{0}(\mathbf{k})=\left(\frac{\hbar^2k_x^2}{2m_x}+\frac{\hbar^2k_y^2}{2m_y}+\epsilon\right)\sigma_z +\mu\label{eq:H_BHZ_single}\\
&\hspace{25mm}+\lambda_xk_x \sigma_x s_z + \lambda_yk_y \sigma_y,\nonumber
\end{align}
where $\sigma_i$ and $s_i$ for $i\in\{x,y,z\}$ are Pauli matrices acting in orbital and spin space, respectively. The parameters $m_x$, $m_y$, $\lambda_x$, and $\lambda_y$ are material-specific constants inherent to the BHZ model~\cite{Bernevig2006}. For simplicity, we assume $m_x,m_y,\lambda_x,\lambda_y>0$ in the following. Furthermore, $\epsilon$ describes a relative energy shift between the particle-like ($\sigma=1$) and hole-like ($\sigma=\bar{1}$) bands, and $\mu$ denotes the chemical potential.

We now proceed to describe the full TI bilayer as shown in Fig.~\ref{fig:model}. Relative to each other, the two TI layers are constructed such that edge states of the same spin polarization propagate in opposite directions for opposite layers. We furthermore account for tunneling between the two layers by a simple spin-conserving tunneling element. Finally, the two layers are proximitized by a top and a bottom $s$-wave superconductor such that the phase difference between them is $\pi$. This could, for example, be achieved by placing a layer of randomly oriented magnetic impurities between one of the layers and the neighboring superconductor~\cite{Schrade2015,Vavra2006,Dam2006}. Alternatively, a superconducting loop connecting the two superconductors allows one to tune the phase difference by varying the enclosed magnetic flux~\cite{Ren2018,Fornieri2018,Shabani2016}. The total Hamiltonian can now be written as $H=\frac{1}{2}\sum_{\mathbf{k}}\Psi_\mathbf{k}^\dagger\mathcal{H}(\mathbf{k})\Psi_\mathbf{k}$ in the basis $\Psi_\mathbf{k}=(\phi_\mathbf{k},\phi_{-\mathbf{k}}^\dagger)$ with $\phi_\mathbf{k}= (\psi_{\mathbf{k}111}$, $\psi_{\mathbf{k}11\bar1}$, $\psi_{\mathbf{k}1\bar11}$, $\psi_{\mathbf{k}1\bar1\bar1}$, $\psi_{\mathbf{k}\bar111}$, $\psi_{\mathbf{k}\bar11\bar1}$, $\psi_{\mathbf{k}\bar1\bar11}$, $\psi_{\mathbf{k}\bar1\bar1\bar1})$, where the electron destruction (creation) operator $\psi_{\mathbf{k}\tau\sigma s}$ ($\psi_{\mathbf{k}\tau\sigma s}^\dagger$) now carries an additional subscript $\tau\in\{1,\bar 1\}$ denoting the layer index.
%
The Hamiltonian density is then given by
\begin{align}
\mathcal{H}(\mathbf{k})&=\left(\frac{\hbar^2k_x^2}{2m_x}+\frac{\hbar^2k_y^2}{2m_y}+\epsilon\right)\eta_z\sigma_z +\mu\eta_z\label{eq:H_BHZ} \\&\quad+\lambda_xk_x \tau_z\sigma_x s_z+ \lambda_yk_y\eta_z \sigma_y+\Delta_{sc}\eta_y\tau_zs_y+\Gamma\eta_z\tau_x,\nonumber
\end{align}
where we have introduced additional Pauli matrices $\tau_i$ and $\eta_i$ for $i\in\{x,y,z\}$ acting in layer and particle-hole space, respectively. The strength of the proximity-induced superconductivity is denoted by $\Delta_{sc}$, while $\Gamma$ denotes the strength of the interlayer tunneling.
%

The Hamiltonian given in Eq.~(\ref{eq:H_BHZ}) is time-reversal symmetric with $\mathcal{T}=is_y\mathcal{K}$ and particle-hole symmetric with $\mathcal{P}=\eta_x\mathcal{K}$, where $\mathcal{K}$ denotes the complex conjugation. As such, our model belongs to the symmetry class DIII~\cite{Ryu2010}. Furthermore, our model has a twofold rotational symmetry around the $z$ axis given by $U_\pi=e^{i\pi\eta_zs_z\sigma_z/2}$. Note that even in the isotropic case $m_x=m_y=m$ and $\lambda_x=\lambda_y=\lambda$, the presence of $\tau_z$ in the term proportional to $\lambda_x$ breaks the usual fourfold rotational symmetry of the BHZ model given by $U_{\pi/2}=e^{i\pi\eta_z s_z(2\sigma_0-\sigma_z)/4}$. 
This will turn out to be crucial to realize the second-order phase proposed in the following. 
However, for $\Gamma=0$, we can define a generalized fourfold rotational symmetry $U'_{\pi/2}=e^{i\pi\eta_z\tau_zs_z(2\sigma_0-\sigma_z)/4}$ such that $U'_{\pi/2}\mathcal{H}(k_x,k_y)[U'_{\pi/2}]^{-1}=\mathcal{H}(-k_y,k_x)$ in the isotropic case.

\begin{figure*}[h!bt]
	\centering
	\includegraphics[width=\textwidth]{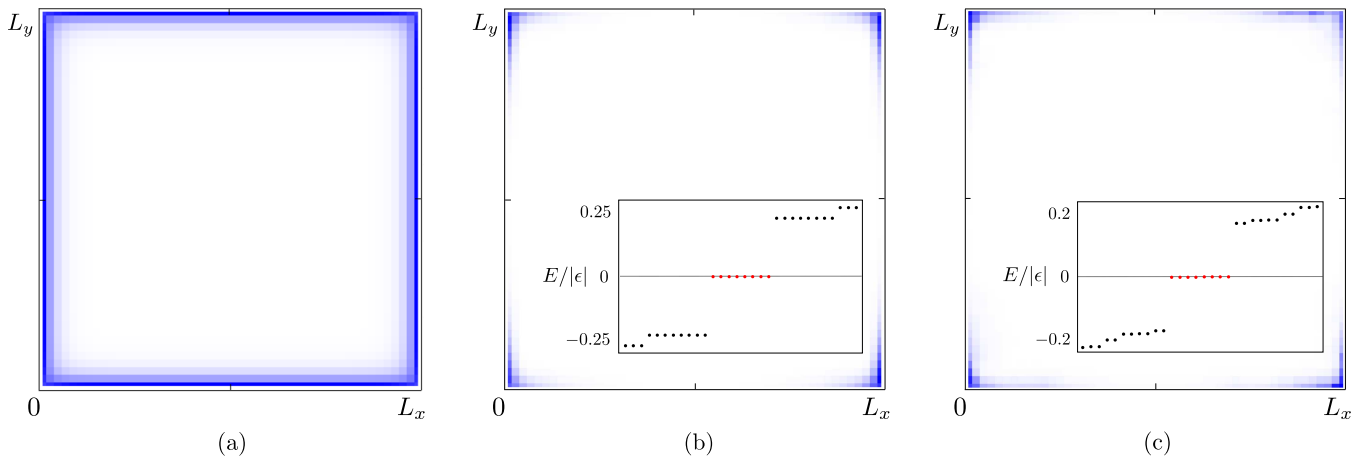}
	\caption{Probability density of low-energy states obtained numerically from a discretized version of Eq.~(\ref{eq:H_BHZ}), see Eq.~(\ref{eq:H_BHZ_discretized}). (a) For $\Delta_{sc}=\Gamma=0$, we find gapless  edge states running along the edges of a large but finite sample. In this case, our model simply corresponds to two decoupled 2D TIs with one Kramers pair of counterpropagating edge states per layer. (b) A Kramers pair of MCSs is localized at each of the four corners of the system for $|\Gamma|>|\Delta_{sc}|$. The inset demonstrates that the energies of these states (red dots) are indeed at zero. Here, we use $\Delta_{sc}/|\epsilon|\approx0.31$ and $\Gamma/|\epsilon|\approx0.63$. (c) The corner states are robust against potential disorder as long as the edge gap remains open. Here, we take the local fluctuations in the chemical potential to follow a normal distribution centered around the mean value $\mu=0$ with standard deviation $\bar\sigma_\mu/|\epsilon|\approx1.13$.
The numerical lattice parameters are $t_x/|\epsilon|=t_y/|\epsilon|=1.25$, $\alpha_x/|\epsilon|=\alpha_y/|\epsilon|=1$, and $L_x=L_y=50$ sites, with the definition of the discretized model and its parameters given in Appendix~\ref{app:lattice}.
 }
	\label{fig:majoranas}
\end{figure*}

\section{Edge states in the first-order phase}
\label{sec:TI}

Let us first consider the case $\Delta_{sc}=\Gamma=0$. In this case, our model simply corresponds to two decoupled copies of the BHZ model. Furthermore, we set $\mu=0$ to simplify our analysis. The bulk spectrum is then given by
\begin{equation} 
E_\pm(\mathbf{k})=\pm\sqrt{\left(\frac{\hbar^2k_x^2}{2m_x}+\frac{\hbar^2k_y^2}{2m_y}+\epsilon\right)^2+\lambda_x^2k_x^2+\lambda_y^2k_y^2}.
\end{equation}
We find that the bulk gap closes at $\mathbf{k}=0$ for $\epsilon=0$, separating a trivial phase for $\epsilon>0$ from a topologically non-trivial TI phase for $\epsilon<0$~\cite{Bernevig2006}. In our case, the latter is characterized by the presence of one pair of counterpropagating helical edge states per layer.

The explicit form of these edge states is readily obtained by following the standard procedure of matching decaying eigenfunctions. Let us first focus on the edges along the $x$ direction. For this, we consider a semi-infinite geometry such that the sample is finite along the $y$ direction and infinite along the $x$ direction. In this setting, $k_x$ remains a good quantum number, while we replace $k_y$ with $-i\partial_y$. For simplicity, we begin by solving for zero-energy eigenstates at $k_x=0$ before perturbatively including linear contributions in $k_x$. Thus, we solve
\begin{equation}
\mathcal{H}(0,-i\partial_y)=\left(\epsilon-\frac{\hbar^2\partial_y^2}{2m_y}\right)\eta_z\sigma_z-i\lambda_y\partial_y\eta_z\sigma_y
\label{eq:Hx}
\end{equation}
for exponentially decaying eigenfunctions $\Phi(y)$ with vanishing boundary conditions $\Phi(y=0)=0$.  As both the layer index as well as the spin-projection along the $z$ axis are good quantum numbers, we can express our solutions as eigenstates of $\tau_z$ and $s_z$. Furthermore, since $\{\mathcal{H}(0,-i\partial_y),\sigma_x\}=0$ and we are looking for zero-energy eigenstates, the solutions are also eigenstates of $\sigma_x$. Therefore, we can write the solutions in terms of eigenstates $|\tau,s,a\rangle$ defined via $\tau_z s_z\sigma_x|\tau,s,a\rangle=\tau sa|\tau,s,a\rangle$, where $a\in\{1,\bar 1\}$ is used to denote the eigenvalue of $\sigma_x$. Explicitly, we find that the solutions are given by
\begin{equation}
\Phi^x_{\tau s}(y)=| \tau,s,1\rangle(e^{-y/\xi_1}-e^{-y/\xi_2})
\label{eq:solx}
\end{equation}
 %
%
with $\xi_{1/2}=(-\lambda_y\pm\sqrt{\beta_y})/(2\epsilon)$ for $\beta_y=\lambda_y^2+2\hbar^2\epsilon/m_y$ and where we have suppressed a normalization factor. Note that since $\epsilon<0$ in the topologically non-trivial phase, we have $\mathrm{Re}(\xi_{1/2})>0$, confirming that our solutions are indeed exponentially localized to the edge of the system.
Furthermore, it is straightforward to see that the solutions are related by time-reversal symmetry as $\mathcal{T}\Phi^x_{\tau s}(y)=\bar{s}\Phi^x_{\tau \bar s}(y)$.

For the edges along the $y$ direction, a similar consideration yields 
\begin{equation}
\mathcal{H}(-i\partial_x,0)=\left(\epsilon-\frac{\hbar^2\partial_x^2}{2m_x}\right)\eta_z\sigma_z-i\lambda_x\partial_x\tau_z s_z\sigma_x.
\label{eq:Hy}
\end{equation}
In this case, the solutions for the edge states turn out to be eigenstates of $\tau_z$, $s_z$, and $\sigma_y$. Therefore, we will write them in terms of eigenstates $|\tau,s,b\rangle$ defined via $\tau_z s_z\sigma_y|\tau,s,b\rangle=\tau sb|\tau,s,b\rangle$, where $b\in\{1,\bar 1\}$ is used to denote the eigenvalue of $\sigma_y$. We arrive at
\begin{equation}
\Phi^y_{\tau s}(x)=| \tau,s,\tau s\rangle(e^{-x/\xi_1'}-e^{-x/\xi_2'})
\label{eq:soly}
\end{equation}
%
with $\xi_{1/2}'=(-\lambda_x\pm\sqrt{\beta_x})/(2\epsilon)$ for $\beta_x=\lambda_x^2+2\hbar^2\epsilon/m_x$ and where we have again omitted a normalization factor.

Finally, the kinetic term governing the low-energy spectrum can be found by taking into account the linear terms in $k_x$ or $k_y$, respectively. Along the $x$ direction, we find that 
\begin{equation}
\lambda_xk_x\langle\Phi^x_{\tau s}|\tau_z\sigma_xs_z|\Phi^x_{\tau's'}\rangle=\tau s\lambda_xk_x\delta_{\tau\tau'}\delta_{ss'}.
\end{equation}
Indeed, the structure of the edge states given in Eq.~(\ref{eq:solx}) makes it immediately clear that states with $\tau s=+1$ ($\tau s=-1$) propagate in the positive (negative) $x$ direction. Similarly, we find that
\begin{equation}
\lambda_yk_y\langle\Phi^y_{\tau s}|\sigma_y|\Phi^y_{\tau's'}\rangle=\tau s\lambda_yk_y\delta_{\tau\tau'}\delta_{ss'}
\end{equation}
along the $y$ direction. Again, states with $\tau s=+1$ ($\tau s=-1$) propagate in the positive (negative) $y$ direction. As expected, we therefore find a pair of counterpropagating gapless edge states per layer, see also Fig.~\ref{fig:majoranas}(a) for a numerical verification. Within each layer, counterpropagating edge states carry opposite spin projections, while counterpropagating edge states in opposite layers carry the same spin projection.

\section{Kramers pairs of Majorana corner states}
\label{sec:cornerstates}

In the following, we take into account the effects of superconductivity and interlayer tunneling in a perturbative way. For this, we assume $\Delta_{sc}$ and $\Gamma$ to be small compared to the bulk gap of the first-order phase, such that their only effect will be to potentially gap out the edge states found above. In order to understand the emergence of corner states, we derive an effective Hamiltonian describing the low-energy edge physics for each edge.

Let us start by considering the tunneling term $\mathcal{H}_\Gamma=\Gamma\eta_z\tau_x$, while keeping $\Delta_{sc}=0$ for the moment. For the edge states along the $x$ direction, we obtain by direct calculation 
\begin{equation}
\langle\Phi^x_{\tau s}|\mathcal{H}_\Gamma|\Phi^x_{\tau's'}\rangle=\Gamma\delta_{\bar\tau\tau'}\delta_{ss'}.
\end{equation}
As such, the tunneling term fully gaps out the edge states along the $x$ direction~\cite{note}. Along the $y$ direction, however, we obtain
\begin{equation}
\langle\Phi^y_{\tau s}|\mathcal{H}_\Gamma|\Phi^y_{\tau's'}\rangle=0
\label{eq:tun_y}
\end{equation}
for all $\tau$, $\tau'$, $s$, and $s'$, which may seem surprising at first. However, this is a direct consequence of the symmetries of the system. Indeed, we note that the system has an additional symmetry $\mathcal{O}=\tau_z\sigma_y$ that anticommutes with the Hamiltonian given in Eq.~(\ref{eq:Hy}). Furthermore, we find $\mathcal{O}|\Phi_{\tau s}^y\rangle=s|\Phi_{\tau s}^y\rangle$. Together with $\{\mathcal{H}_\mathrm{tun},\mathcal{O}\}=0$, we then find $\langle\Phi^y_{\tau s}|\mathcal{H}_\Gamma|\Phi^y_{\bar\tau s}\rangle=-\langle\Phi^y_{\tau s}|\mathcal{H}_\Gamma|\Phi^y_{\bar\tau s}\rangle=0$. The other matrix elements are trivially zero by the definition of $\mathcal{H}_{\Gamma}$, which confirms Eq.~(\ref{eq:tun_y}).

Let us now additionally consider the effect of superconductivity. Clearly, superconductivity will open a gap along all edges, leading to an effective edge Hamiltonian of the form
\begin{equation}
H_{\mathrm{eff}}^x(k_x)=\lambda_xk_x\tau_z s_z+\Gamma\eta_z\tau_x+\Delta_{sc}\eta_y\tau_zs_y
\label{eq:Heff_x}
\end{equation}
for the edges along the $x$ direction and
\begin{equation}
H_\mathrm{eff}^y(k_y)=\lambda_yk_y \tau_zs_z+\Delta_{sc}\eta_y\tau_zs_y
\label{eq:Heff_y}
\end{equation}
for the edges along the $y$ direction. From this it becomes clear that as long as $|\Delta_{sc}|>0$, the edges along the $y$ direction are trivially gapped by superconductivity. Along the $x$ direction, on the other hand, the edge gap closes at $|\Delta_{sc}|=|\Gamma|$. Indeed, we recognize Eq.~(\ref{eq:Heff_x}) to be the Hamiltonian of a time-reversal invariant 1D TSC as discussed in Ref.~\cite{Keselman2013}. This system hosts a Kramers pair of Majorana bound states at domain walls separating a topological phase with $|\Gamma|>|\Delta_{sc}|$ from a trivial phase. In our model, these domain walls appear at the corners between $x$ and $y$ edges, leaving us with a Kramers pair of MCSs at all four corners of a rectangular sample. In Fig.~\ref{fig:majoranas}(b), we have verified the existence of the corner states numerically. Furthermore, we have tested the stability of the corner states against potential disorder, see Fig.~\ref{fig:majoranas}(c). In particular, we note that the symmetry $\mathcal{O}$ used to derive the corner states can be broken as long as the edge gap remains open. Indeed, the MCSs are protected solely by particle-hole and time-reversal symmetry and do not rely on any additional spatial symmetry.

\section{Unequal tunneling amplitudes for particle-like and hole-like bands}
\label{sec:exp}

In realistic setups we generally expect the interlayer tunneling amplitude for the particle-like and hole-like bands to be different in size. This constitutes an obstruction to the second-order topological phase presented here. In the following, we account for this by introducing a refined tunneling Hamiltonian
\begin{equation}
\mathcal{H}_\Gamma=\frac{\Gamma_e+\Gamma_h}{2}\eta_z\tau_x+\frac{\Gamma_e-\Gamma_h}{2}\eta_z\tau_x\sigma_z,
\end{equation}
where $\Gamma_e$ ($\Gamma_h$) is used to denote the tunneling amplitude for electrons (holes). Calculating the effective Hamiltonian along the $x$ and $y$ direction using the edge state solutions given in Eqs.~(\ref{eq:solx}) and (\ref{eq:soly}), we find
\begin{equation}
H_{\mathrm{eff}}^x(k_x)=\lambda_xk_x\tau_z s_z+\frac{\Gamma_e+\Gamma_h}{2}\eta_z\tau_x+\Delta_{sc}\eta_y\tau_zs_y
\end{equation}
for the edges along the $x$ direction and
\begin{equation}
H_\mathrm{eff}^y(k_y)=\lambda_yk_y \tau_zs_z+\frac{\Gamma_e-\Gamma_h}{2}\eta_z\tau_x+\Delta_{sc}\eta_y\tau_zs_y
\end{equation}
for the edges along the $y$ direction. 
We therefore find that the SOTSC phase persists if $|\Gamma_e-\Gamma_h|<2|\Delta_{sc}|<|\Gamma_e+\Gamma_h|$ or $|\Gamma_e+\Gamma_h|<2|\Delta_{sc}|<|\Gamma_e-\Gamma_h|$.

Studies of HgTe double quantum well structures have estimated $\Gamma_h$ to be negligibly small compared to $\Gamma_e$ in the experimentally accessible parameter range, i.e., $\Gamma_h\approx 0$~\cite{Michetti2012,Michetti2013,Krishtopenko2016}. This excludes the double-well setup as a possible realization of the topological phase proposed here. However, other systems with similar low-energy properties may circumvent this issue. In particular, mono- and few-layer $\mathrm{Fe}(\mathrm{Te}_{1-x}\mathrm{Se}_x)$ have recently been claimed to exhibit a low-energy band structure described by the BHZ Hamiltonian~\cite{Wu2016}. Similarly, the 2D transition metal dichalchogenides (TMDCs) $\mathrm{MX}_2$ with $\mathrm{M}\in\{\mathrm{W},\mathrm{Mo}\}$ and $\mathrm{X}\in\{\mathrm{S},\mathrm{Se},\mathrm{Te}\}$ have been shown to exhibit the desired low-energy effective band structure~\cite{Qian2014}. It would therefore be interesting to investigate TI bilayers built from these materials as potential experimental realizations of the SOTSC proposed in this work. 

\begin{figure*}[thb]
	\centering
	\includegraphics[width=\textwidth]{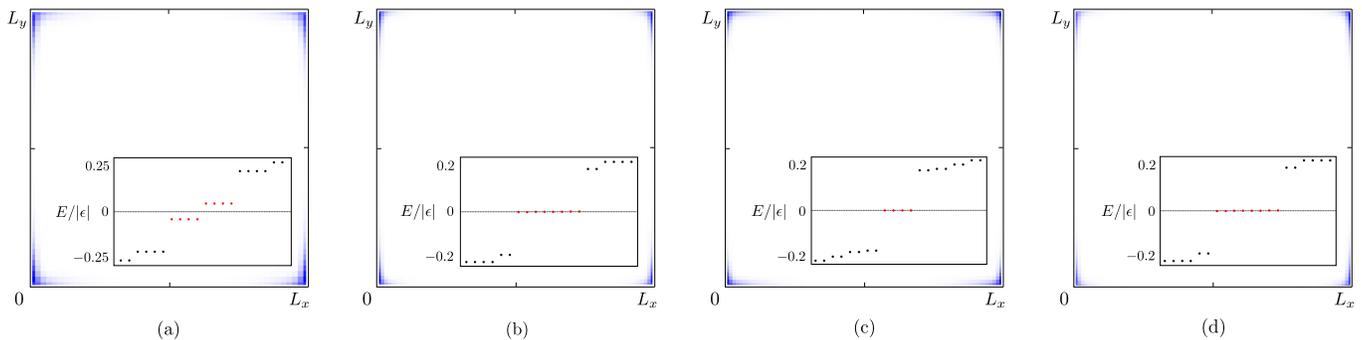}
	\caption{(a) Probability density of low-energy states obtained numerically from a discretized version of Eq.~(\ref{eq:H_BHZ}) for a sample of $L_x=L_y=50$ sites with $\Delta_{sc}/|\epsilon|\approx0.31$ and $\Gamma/|\epsilon|\approx0.63$ and in the additional presence of an out-of-plane Zeeman field of strength $\Delta_{Z,\perp}/|\epsilon|\approx0.04$. We find that the two Kramers partners of MCSs at each corner hybridize and split away from zero energy, see the red dots in the inset. (b)-(d) Probability density of the lowest-energy state obtained numerically from a discretized version of Eq.~(\ref{eq:H_BHZ}) for a sample of $L_x=L_y=80$ sites in the additional presence of an in-plane Zeeman field oriented along the $x$ direction. (b) For a weak Zeeman field $0\leq\Delta_{Z,||}<\Gamma-\Delta_{sc}$, there are two MCSs per corner. Here, we have used $\Delta_{sc}/|\epsilon|=0.25$, $\Gamma/|\epsilon|=0.5$ and $\Delta_{Z,||}/|\epsilon|\approx0.13$. (c) In the intermediate regime $|\Gamma-\Delta_{sc}|<\Delta_{Z,||}<\Gamma+\Delta_{sc}$, we find one MCS per corner. Here, we have used $\Delta_{sc}/|\epsilon|=\Gamma/|\epsilon|\approx0.38$ and $\Delta_{Z,||}/|\epsilon|\approx0.19$. (d) For strong Zeeman fields  $\Gamma+\Delta_{sc}<\Delta_{Z,||}$, we again find two MCSs per corner. Here, we have used $\Delta_{sc}/|\epsilon|=0.25$, $\Gamma/|\epsilon|\approx0.13$ and $\Delta_{Z,||}/|\epsilon|=0.5$. The other numerical parameters are the same as in Fig.~\ref{fig:majoranas}.}
	\label{fig:in_plane}
\end{figure*}

\section{Effect of Zeeman field and single-MCS phase}
\label{sec:single_MCS}

In this section, we additionally comment on the effects of a Zeeman field, which we again assume to be sufficiently weak  compared to the bulk gap of the first-order phase. Since time-reversal symmetry is now broken, the fate of the MCSs is not a priori clear in this case. Indeed, we find that in the presence of an out-of-plane Zeeman term $\mathcal{H}_{Z,\perp}=\Delta_{Z,\perp}\eta_zs_z$, the Kramers pairs at each corner hybridize and split away from zero energy, see Fig.~\ref{fig:in_plane}(a). Thus, the topological phase is destroyed in this case. On the other hand, however, we find that an in-plane Zeeman field does not completely destroy the topological properties of the system, but instead leads to a much richer phase diagram exhibiting two nonequivalent regions with two MCSs per corner as well as an intermediate region with just one MCS per corner. For concreteness, let us consider the case when the in-plane field is oriented along the $x$ direction, i.e., $\mathcal{H}_{Z,||}=\Delta_{Z,||}\eta_zs_x$. Other orientations of the in-plane Zeeman field lead to qualitatively identical results. Note that here we take the Zeeman field to be of equal strength for both the particle-like and the hole-like bands. Depending on the experimental realization, one may again need to generalize this in a way similar to the treatment of unequal tunneling amplitudes in Sec.~\ref{sec:exp}. Calculating again the effective edge Hamiltonian, we find that
\begin{equation}
H_{\mathrm{eff}}^x(k_x)=\lambda_xk_x\tau_z s_z+\Gamma\eta_z\tau_x+\Delta_{sc}\eta_y\tau_zs_y+\Delta_{Z,||}\eta_zs_x,
\end{equation}
while the effective edge Hamiltonian along the $y$ direction is still given by Eq.~(\ref{eq:Heff_y}). Therefore, the edges along the $y$ direction remain trivially gapped by superconductivity, whereas the edge gap along the $x$ direction now closes at $\Delta_{Z,||}+\Gamma=\pm\Delta_{sc}$ and $\Delta_{Z,||}-\Gamma=\pm\Delta_{sc}$.

\begin{figure*}[ht]
	\centering
	\includegraphics[width=0.8\textwidth]{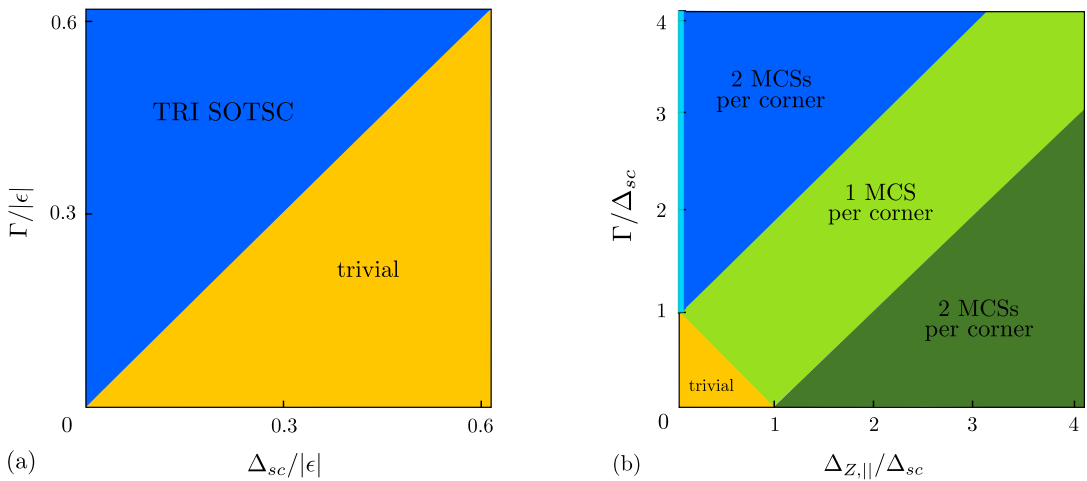}
	\caption{(a) Phase diagram of the time-reversal invariant system discussed in Sec.~\ref{sec:cornerstates} as a function of $\Gamma$ and $\Delta_{sc}$. Note that we focus on the regime where both $\Gamma$ and $\Delta_{sc}$ are sufficiently small compared to the bulk gap of the first-order phase 
		such that our pertubative treatment of these terms is justified. We find that the edge gap closes and reopens for $\Gamma=\Delta_{sc}$, corresponding to the phase transition between the topologically trivial phase (shaded in yellow) and the SOTSC phase with a Kramers pair of MCSs per corner (shaded in dark blue). (b) In the presence of an in-plane Zeeman field, the phase diagram becomes even richer. For a weak Zeeman field $0\leq\Delta_{Z,||}<\Gamma-\Delta_{sc}$, there are two MCSs per corner. The light blue line with $\Delta_{Z,||}=0$ denotes the time-reversal invariant SOTSC phase discussed in Sec.~\ref{sec:cornerstates}, where the two MCSs at each corner are protected by time-reversal symmetry. For $0<\Delta_{Z,||}<\Gamma-\Delta_{sc}$, see the region shaded in dark blue, the two MCSs per corner remain intact. However, they are now no longer protected by time-reversal symmetry and may split away from zero energy in the presence of disorder. In the intermediate regime $|\Gamma-\Delta_{sc}|<\Delta_{Z,||}<\Gamma+\Delta_{sc}$, see the region shaded in light green, we instead find that there is only one MCS per corner. Most interestingly, we find that for $\Gamma=\Delta_{sc}$ even an infinitesimal Zeeman field can drive the system into a SOTSC phase with one MCS per corner, as the competing tunneling and superconducting terms completely cancel each other. Finally, for $\Gamma+\Delta_{sc}<\Delta_{Z,||}$ (shaded in dark green), we again find two MCSs per corner. Indeed, this regime can be connected to the limit $\Gamma=0$ and $\Delta_{sc}<\Delta_{Z,||}$. In this case, we simply deal with two independent copies of the system proposed in Ref.~\cite{Wu2019}.}
	\label{fig:phase_diagram}
\end{figure*}

In the following, we comment on the different (second-order) topological phases separated by the above gap closing lines. For simplicity, we assume that $\Delta_{sc},\Delta_{Z,||},\Gamma\geq0$. Firstly, we have checked numerically that for $0<\Delta_{Z,||}<\Gamma-\Delta_{sc}$, the two MCSs per corner remain intact, see Fig.~\ref{fig:in_plane}(b). However, they are now no longer protected by time-reversal symmetry and may split away from zero energy in the presence of magnetic disorder~\cite{Hsu2015,Reeg2017,Hoffman2016,olesia}. Secondly, we find that in the intermediate regime $|\Gamma-\Delta_{sc}|<\Delta_{Z,||}<\Gamma+\Delta_{sc}$, there is only one MCS per corner, see Fig.~\ref{fig:in_plane}(c). Most interestingly, we find that for $\Gamma=\Delta_{sc}$ even an infinitesimal Zeeman field can drive the system into a SOTSC phase with one MCS per corner, as the competing tunneling and superconducting terms completely cancel each other. Finally, for $\Gamma+\Delta_{sc}<\Delta_{Z,||}$, we again find two MCSs per corner, see Fig.~\ref{fig:in_plane}(d). Indeed, this regime is in the same region of the phase diagram as the limit $\Gamma=0$ and $\Delta_{sc}<\Delta_{Z,||}$. In this case, we simply deal with two decoupled TI layers subjected to an in-plane Zeeman field. Indeed, a single TI layer in the presence of an in-plane Zeeman field has been shown to exhibit a SOTSC phase in Ref.~\cite{Wu2019}. To summarize, Fig.~\ref{fig:phase_diagram} displays the phase diagram of our system both in the absence [Fig.~\ref{fig:phase_diagram}(a)] and presence [Fig.~\ref{fig:phase_diagram}(b)] of an in-plane Zeeman field. 

\begin{figure}[t]
	\centering
	\includegraphics[width=0.9\columnwidth]{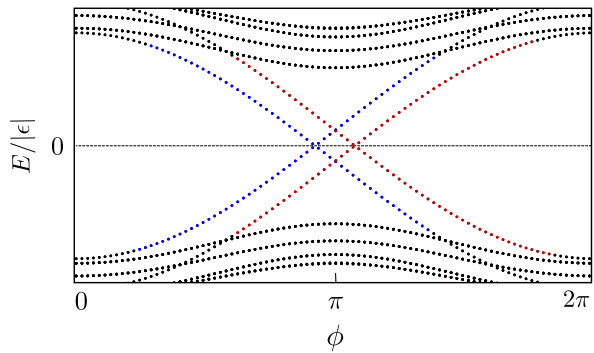}
	\caption{Energy spectrum obtained numerically from a discretized version of Eq.~(\ref{eq:H_BHZ}) for a sample of $L_x=L_y=50$ sites in the presence of an out-of-plane Zeeman field of strength $\Delta_{Z,\perp}/|\epsilon|\approx0.04$ and for a varying superconducting phase difference $\phi$. The finite out-of-plane Zeeman field shifts the corner states away from zero energy at $\phi=\pi$, see also Fig.~\ref{fig:in_plane}(a). However, by changing the superconducting phase away from $\phi=\pi$, the corner states at a given corner can be brought back to zero energy. The red (blue) lines are two-fold degenerate and correspond to the upper right and lower left (upper left and lower right) corners. The other numerical parameters are the same as in Fig.~\ref{fig:majoranas}.}
	\label{fig:phase_difference}
\end{figure}

It is worth noting that another way to break time-reversal symmetry is by detuning the superconducting phase difference away from $\pi$. In this case, the superconducting term entering in Eq.~(\ref{eq:H_BHZ}) takes the more general form
\begin{equation}
\mathcal{H}_{sc}=\frac{\Delta_{sc}}{2}[(1+\tau_z)\eta_ys_y+(1-\tau_z)(\mathrm{cos}\phi\,\eta_ys_y-\mathrm{sin}\phi\,\eta_xs_y)],
\end{equation}
where $\phi=\pi$ reproduces the time-reversal invariant case discussed above. For deviations from $\phi=\pi$, we find that the MCSs at each corner hybridize and split away from zero energy. However, similarly as in Ref.~\cite{Keselman2013}, we can exploit the interplay between a detuning of the superconducting phase difference and an out-of-plane Zeeman field to bring the corner states at a given corner back to zero energy. This is illustrated in Fig.~\ref{fig:phase_difference}. In particular, we note that this mechanism provides us with a way to go from a phase with zero-energy corner states at all four corner of the sample to a phase with zero-energy corner states only at two opposite corners of the sample.

Finally, let us mention that also the fermion parity pumping effect discussed in Ref.~\cite{Keselman2013} can be observed in our system when the superconducting phase difference is adiabatically varied from $0$ to $2\pi$. This follows immediately from the fact that, in the low-energy limit, our system is nothing but two copies of the 1D system considered in Ref.~\cite{Keselman2013} separated by two topologically trivial regions.

For the single-MCS phase, on the other hand, we find that the MCSs persist for a finite range of phase detunings up to a critical value at which the edge gaps close.

\section{Conclusions}
\label{sec:conclusions}

We have proposed a versatile and experimentally feasible platform that realizes a 2D time-reversal invariant SOTSC phase hosting Kramers pairs of MCSs. Our setup consists of two tunnel-coupled 2D TIs proximitized by a top and bottom $s$-wave superconductor with a phase difference of $\pi$ between them. In the regime where the interlayer tunneling dominates over the proximity-induced superconductivity, we find a Kramers pair of MCSs at all four corners of a rectangular sample. Additionally, we have shown that a weak but finite in-plane Zeeman field further enriches the phase diagram. In particular, we find that there are now two nonequivalent SOTSC phases with two MCSs per corner as well as an intermediate phase with just one MCS per corner. Most interestingly, this single-MCS phase is accessible even for very weak Zeeman fields if the tunneling and superconducting term are of comparable strength. As the requirement of a strong magnetic field in combination with superconductivity is experimentally problematic, our proposal constitutes an interesting alternative route towards the realization of a single-MCS SOTSC phase that is accessible even for weak magnetic fields.

\acknowledgements

This work was supported by the Swiss National Science Foundation and NCCR QSIT. This project received funding from the European Union’s Horizon 2020 research and innovation program (ERC Starting Grant, grant agreement No 757725).

\appendix

\section{Lattice model for the BHZ Hamiltonian}
\label{app:lattice}

In this Appendix, we present the discretized version of the Hamiltonian given in Eq.~(\ref{eq:H_BHZ}). In momentum space, the discretized Hamiltonian reads
\begin{align}
&\mathcal{H}(\mathbf{k})=[-2t_x\mathrm{cos}(k_xa_x)-2t_y\mathrm{cos}(k_ya_y)]\eta_z \sigma_z\nonumber \\&+(\epsilon+2t_x+2t_y)\eta_z\sigma_z+\mu\eta_z+2\alpha_x\mathrm{sin}(k_xa_x) \tau_z\sigma_x s_z\nonumber\\&+ 2\alpha_y\mathrm{sin}(k_ya_y)\eta_z \sigma_y+\Delta_{sc}\eta_y\tau_zs_y+\Gamma\eta_z\tau_x.\label{eq:H_BHZ_discretized}
\end{align}
Here, $a_x$ ($a_y$) is the lattice spacing along the $x$ ($y$) direction. The spin-conserving hopping amplitude $t_x$ ($t_y$) defines the effective mass along the $x$ ($y$) direction via $t_x=\hbar^2/(2m_xa_x^2)$ [$t_y=\hbar^2/(2m_ya_y^2)$]. Similarly, $\alpha_x$ ($\alpha_y$) is related to $\lambda_x$ ($\lambda_y$) via $\alpha_x=\lambda_x/(2a_x)$ [$\alpha_y=\lambda_y/(2a_y$)]~\cite{Volpez2019b}. Note that in the main part of our work, we focus on the isotropic case $t_x=t_y=t$ and $\alpha_x=\alpha_y=\alpha$. The strongly anisotropic case could, for example, be realized in the coupled-wire approach~\cite{fg1,fg2,fg3,Meng2015,Teo2014,Klinovaja2014c,Klinovaja2014d,Sagi2014,Sagi2015,Neupert2014,Klinovaja2015,Meng2015b}.

\bibliographystyle{unsrt}

\begin{thebibliography}{}
\bibitem{comment}
	We note that while Eq.~(1) presents a rather simple toy model, the main conclusions of the following sections -- in particular regarding the existence of the MCSs -- remain unaltered if more elaborate corrections to the BHZ Hamiltonian are taken into account as long as these remain small enough compared to the other energy scales in the system.
\bibitem{note}
	The twofold rotational symmetry of the model ensures that this is indeed the case for \emph{both} edges along the $x$ direction.
\end{thebibliography}

\end{document}